# New approach for solving master equations of density operator for the Jaynes Cummings Model with Cavity Damping

Seyed Mahmoud Ashrafi[1] and Mohammad Reza Bazrafkan

Physics Department, Faculty of Science, I. K. I. University, Qazvin, Iran

By introducing thermo- entangled state representation $|\eta\rangle$, which can map master equations of density operator in quantum statistics as state-vector evolution equations, and using ''dissipative interaction picture'' we solve the master equation of Jaynes- Cummings model with cavity damping. In addition we derive the Wigner function for density operator when the atom is initially in the up state $|\uparrow\rangle$ and the cavity mode is in coherent state.

## 1. Introduction

As is well known, decoherence dynamics and the various methods for controlling this process is one of the important topics in quantum optics and the theory of quantum open systems can help for better understanding of how quantum decoherence is processed. In nature most quantum systems are immersed in reservoir that this systems introduce open system [1, 2] the time evolution of an open quantum system interacting with a memory-less environment can be expressed as a Lindblad master equation as follows[3, 4]

$$\frac{d\hat{\rho}(t)}{dt} = -\frac{i}{\hbar}[\hat{H},\hat{\rho}] + \sum_n \left[ \hat{L}_n \hat{\rho}(t) \hat{L}_n^\dagger - \frac{1}{2}\hat{L}_n^\dagger \hat{L}_n \hat{\rho}(t) - \frac{1}{2}\hat{\rho}(t)\hat{L}_n^\dagger \hat{L}_n \right]. \tag{1}$$

Where $\hat{\rho}(t)$ is, the density of the system, $\hat{H}$ is the Hamiltonian of the system and $\hat{L}_n$ is the so-called Lindblad operators that are supposed to model the effect of the environment on the system. Equation (1) describes the general case of a quantum open system. Using some quantum –mechanical models for the reservoir, one can find a differential equation for the time evolution of reduced density operator of the system.

Let us consider a two-level system interact with a resonant mode of the electromagnetic field, Jaynes-Coumming model [5, 6], this model is described without the rotating wave approximation by the Hamiltonian ($\hbar = 1$)

$$\hat{H} = \omega \hat{a}^\dagger \hat{a} + \lambda(\hat{a}^\dagger + \hat{a})(\hat{\sigma}_+ + \hat{\sigma}_-). \tag{2}$$

If damping is contained in J-C model, one has to solve

$$\frac{d\hat{\rho}(t)}{dt} = -i[\hat{H},\hat{\rho}] + \kappa L_{ir}[\hat{\rho}], \tag{3}$$

Where $L_{ir}[\rho]$ is for $T = 0$

$$L_{ir}[\hat{\rho}] = \kappa(2\hat{a}\hat{\rho}\hat{a}^\dagger - \hat{a}^\dagger \hat{a}\hat{\rho} - \hat{\rho}\hat{a}^\dagger \hat{a}). \tag{4}$$

There is a newly developed method, introduced by Fan and Fan, [7] from which we can map the density operator by a vector of two-mode Fock space whose first mode is the system mode and the second mode is a fictitious one, and consequently the master equation appears as a Schrodinger- like time evolution equation.

---

[**] Corresponding author: ashrafi@ikiu.ac.ir

In this present paper, instead of using phase space representation of density operator [8,9] we adopted thermo- entangled state (TES) representation to treat time evolution of density operator in the J-C model with cavity damping where Xu and Yuan derived density operator for a Raman- coupled model in cavity damping [10]. This paper is organized as follows

In sect.2, we review the (TES) representation and in sect.3, the equation of motion of the density operator is decoupled. In sec.4, we solve the master equation of J-C model by using of "dissipative interaction picture" in entangled representation. In Sec. 5 Wigner function is derived for a special initial condition. Section 6 is devoted to calculation. In the Appendix, we derive a used relation in the text.

## 2. Brief Review of the Thermo Entangled State Representation

In this section, we first briefly review the TES and construct the TES representation in the doubled Fock space in the following form [11-14]

$$|\eta\rangle = \exp\left(-\frac{1}{2}|\eta|^2 + \eta\hat{a}^\dagger - \eta^*\hat{b}^\dagger + \hat{a}^\dagger\hat{b}^\dagger\right)|0,\tilde{0}\rangle, \qquad \eta \in \mathbb{C}, \tag{5}$$

Where $(\hat{a},\hat{a}^\dagger)$ and $(\hat{b},\hat{b}^\dagger)$ are the pairs of canonical annihilation and creation operators which the first (physical) and the second (fictitious) mode respectively. Defining $\hat{\eta} \equiv \hat{a} - \hat{b}^\dagger$ and $\hat{\xi} = \hat{a} + \hat{b}^\dagger$, one can recognize that Eq. (5) is eigen-vector of.

$$\begin{aligned}\hat{\eta}|\eta\rangle &= \eta|\eta\rangle, \\ \hat{\eta}^\dagger|\eta\rangle &= \eta^*|\eta\rangle.\end{aligned} \tag{6}$$

Obviously, When $\eta = 0$

$$|\eta=0\rangle = \exp\left(-\frac{1}{2}|\eta|^2\right)|0,0\rangle = \sum_{n=0}^{\infty}|n,n\rangle. \tag{7}$$

And it is obtain that

$$\begin{aligned}\hat{a}|\eta=0\rangle &= \hat{b}^\dagger|\eta=0\rangle, \\ \hat{a}^\dagger|\eta=0\rangle &= \hat{b}|\eta=0\rangle.\end{aligned} \tag{8}$$

By means of technique (IWOP), one can prove that makes up a complete and orthonormal set

$$\int \frac{d^2\eta}{\pi}|\eta\rangle\langle\eta| = 1, \qquad \langle\eta'|\eta\rangle = \pi\delta^{(2)}(\eta'-\eta). \tag{9}$$

And, it is convenient to introduce the state $|\rho(t)\rangle$ [15] for density operator

$$|\rho(t)\rangle \equiv \hat{\rho}(t)|\eta=0\rangle. \tag{10}$$

This means that we can uniquely represent any density operator $\hat{\rho}(t)$ of physical mode by vector $|\rho(t)\rangle$ from the two-mode Hilbert space. This vector contains all information on the quantum state as the density operator.

## 3. Time Evolution of Density Operator in the J-C Model

In other to describe Eq. (3) in the interaction picture, which may be called rotational interaction picture, we use of the unitary operator

$$U(t) = \exp(-i\omega t\hat{a}^\dagger\hat{a}). \tag{11}$$

Therefore we have

$$U^{-1}\hat{\rho}_s U \Rightarrow \hat{\rho}, \qquad U^{-1}\hat{H}U \Rightarrow \hat{H}, \qquad (12)$$

Inserting Eq. (12) into Eq. (3) we derive the master equation in the rotational interaction picture as follows

$$\frac{d\hat{\rho}(t)}{dt} = -i\left[\lambda(\hat{a}^\dagger e^{i\omega t} + \hat{a}e^{-i\omega t})(\hat{\sigma}_+ + \hat{\sigma}_-), \hat{\rho}\right] + \kappa L_{ir}[\hat{\rho}] \qquad (13)$$

We write the density operator

$$\hat{\rho} = \frac{1}{2}\sum_{i=0}^{3}\hat{\rho}_i \sigma_i = \begin{pmatrix} \hat{\rho}_{11} & \hat{\rho}_{12} \\ \hat{\rho}_{21} & \hat{\rho}_{22} \end{pmatrix} = \frac{1}{2}\begin{pmatrix} \hat{\rho}_0 + \hat{\rho}_3 & \hat{\rho}_1 - i\hat{\rho}_2 \\ \hat{\rho}_1 + i\hat{\rho}_2 & \hat{\rho}_0 - \hat{\rho}_3 \end{pmatrix} \qquad (14)$$

Using the Pauli spin matrices $\hat{\sigma}_+ + \hat{\sigma}_- = \hat{\sigma}_1, \hat{\sigma}_y = \hat{\sigma}_2, \hat{\sigma}_z = \hat{\sigma}_3$ and the unit matrix $\hat{\sigma}_0 = \hat{1}$, Note that

$$\hat{\rho}_{11} = \langle\uparrow|\hat{\rho}|\uparrow\rangle,\ \hat{\rho}_{12} = \langle\uparrow|\hat{\rho}|\downarrow\rangle,\ \hat{\rho}_{21} = \langle\downarrow|\hat{\rho}|\uparrow\rangle,\ \hat{\rho}_{22} = \langle\downarrow|\hat{\rho}|\downarrow\rangle. \qquad (15)$$

And

$$\begin{aligned}
\hat{\rho}_0 &= \hat{\rho}_{11} + \hat{\rho}_{22}, & \hat{\rho}_1 &= \hat{\rho}_{12} + \hat{\rho}_{21}, \\
\hat{\rho}_2 &= i(\hat{\rho}_{12} - \hat{\rho}_{21}), & \hat{\rho}_3 &= \hat{\rho}_{11} - \hat{\rho}_{22}.
\end{aligned} \qquad (16)$$

According to the above equations, Eq. (13) can be split into the following four equations

$$\begin{aligned}
\dot{\hat{\rho}}_0 &= -i\lambda\left[(\hat{a}^\dagger e^{i\omega t} + \hat{a}e^{-i\omega t}), \hat{\rho}_1\right] + \kappa L_{ir}[\hat{\rho}_0], \\
\dot{\hat{\rho}}_1 &= -i\lambda\left[(\hat{a}^\dagger e^{i\omega t} + \hat{a}e^{-i\omega t}), \hat{\rho}_0\right] + \kappa L_{ir}[\hat{\rho}_1], \\
\dot{\hat{\rho}}_2 &= -\lambda\{(\hat{a}^\dagger e^{i\omega t} + \hat{a}e^{-i\omega t})\hat{\rho}_3 + \hat{\rho}_3(\hat{a}^\dagger e^{i\omega t} + \hat{a}e^{-i\omega t})\} + \kappa L_{ir}[\hat{\rho}_2], \\
\dot{\hat{\rho}}_3 &= \lambda\{(\hat{a}^\dagger e^{i\omega t} + \hat{a}e^{-i\omega t})\hat{\rho}_2 + \hat{\rho}_2(\hat{a}^\dagger e^{i\omega t} + \hat{a}e^{-i\omega t})\} + \kappa L_{ir}[\hat{\rho}_3].
\end{aligned} \qquad (17)$$

One can find that the first two and the last equation are coupled, respectively. The first two equations can be decoupled as

$$\dot{\hat{\rho}}_\pm = \mp i\lambda\left[\hat{a}^\dagger e^{i\omega t} + \hat{a}e^{-i\omega t}, \hat{\rho}_\pm\right] + \kappa L_{ir}[\hat{\rho}_\pm] \qquad (18)$$

Where, we introduced

$$\hat{\rho}_\pm = \hat{\rho}_0 \pm \hat{\rho}_1. \qquad (19)$$

The other two equations can be decoupled. By taking

$$\hat{\rho}_c = \hat{\rho}_3 + i\hat{\rho}_2. \qquad (20)$$

Obviously

$$\dot{\hat{\rho}}_c = -i\lambda\{(\hat{a}^\dagger e^{i\omega t} + \hat{a}e^{-i\omega t})\hat{\rho}_c + \hat{\rho}_c(\hat{a}^\dagger e^{i\omega t} + \hat{a}e^{-i\omega t})\} + \kappa L_{ir}[\hat{\rho}_c]. \qquad (21)$$

It is seen that solving the equation of motion in Eq. (13) is equivalent to solving Eq. (18) and Eq. (21).

## 4. Unraveling Master Equations through Entangled State Representation

In this section, we unravel the master equations Eq. (18) and Eq. (19) through (TES) representation. Therefore acting both hand some Eq. (18) in $|\eta = 0\rangle$ and using Eq. (8), we can find a Schrodinger-like time evolution equation

$$\begin{aligned}
\frac{d|\rho_\pm(t)\rangle}{dt} &= \left\{\mp i\lambda\left[\hat{a}e^{-i\omega t} + \hat{a}^\dagger e^{i\omega t} - \hat{b}e^{i\omega t} - \hat{b}^\dagger e^{-i\omega t}\right] + \frac{\gamma}{2}\left(2\hat{a}\hat{b} - \hat{a}^\dagger \hat{a} - \hat{b}^\dagger \hat{b}\right)\right\}|\rho_\pm(t)\rangle \\
&= \left\{\mp i\lambda(\hat{\eta}e^{-i\omega t} + \hat{\eta}^\dagger e^{i\omega t}) + \frac{\gamma}{2}\hat{L}\right\} \\
&= \left\{\hat{g}_\pm(t) + \frac{\gamma}{2}\hat{L}\right\}|\rho_\pm(t)\rangle,
\end{aligned} \qquad (22)$$

Where

$$\hat{g}_\pm(t) = \mp i\lambda(\hat{\eta}e^{-i\omega t} + \hat{\eta}^\dagger e^{i\omega t}),\ \hat{L} = (2\hat{a}\hat{b} - \hat{a}^\dagger \hat{a} - \hat{b}^\dagger \hat{b}). \qquad (23)$$

By using of commutation relations $\left[\hat{L}, \hat{\eta}\right] = -\hat{\eta}$ and $\left[\hat{L}, \hat{\eta}^\dagger\right] = -\hat{\eta}^\dagger$ one can find that

$$\left[\hat{g}_\pm(t) + \frac{\gamma}{2}\hat{L}, \hat{g}_\pm(t') + \frac{\gamma}{2}\hat{L}\right] \neq 0. \tag{24}$$

It is difficult to solve Eq. (22) therefore we define "dissipative interaction picture" in entangled representation [15, 16] for $\rho_\pm(t)$ as follows

$$\left|\rho_\pm(t)\right\rangle_I = e^{-\frac{\gamma t}{2}\hat{L}}\left|\rho_\pm(t)\right\rangle. \tag{25}$$

For new variable $\left|\rho_\pm(t)\right\rangle_I$ we have

$$\frac{d\left|\rho_\pm(t)\right\rangle_I}{dt} = \hat{g}^I_\pm(t)\left|\rho_\pm(t)\right\rangle_I, \qquad \hat{g}^I_\pm(t) \triangleq e^{-\frac{\gamma t}{2}\hat{L}}\hat{g}_\pm(t)e^{\frac{\gamma t}{2}\hat{L}}. \tag{26}$$

Where

$$\begin{aligned}\hat{g}^I_\pm(t) &= e^{-\frac{\gamma t}{2}\hat{L}}\hat{g}_\pm(t)e^{\frac{\gamma t}{2}\hat{L}} \\ &= \mp i\lambda\, e^{\frac{\gamma t}{2}}\left\{\hat{\eta}e^{-i\omega t} + \hat{\eta}^\dagger e^{i\omega t}\right\}.\end{aligned} \tag{27}$$

Now we have

$$\left[\hat{g}^I_\pm(t), \hat{g}^I_\pm(t')\right] = 0. \tag{28}$$

Therefore we can integrate Eq. (26) and find

$$\left|\rho_\pm(t)\right\rangle_I = \exp\{\int_0^t dt'\hat{g}^I_\pm(t')\}\left|\rho_\pm(0)\right\rangle_s, \tag{29}$$

To find more simplified form of (29) we write

$$\{\int_0^t dt'\hat{g}_I(t')\} = \{\lambda_\pm(t)\hat{\eta}^\dagger - \lambda_\pm^*(t)\hat{\eta}\}, \tag{30}$$

Where the time dependent parameter $\lambda_\pm(t)$ is defined as

$$\lambda_\pm(t) \triangleq \mp i\lambda\int_0^t dt'\, e^{i\omega t' + \frac{\gamma t'}{2}} = \mp\frac{2i\lambda}{2i\omega + \gamma}\{e^{\frac{\gamma t}{2}+i\omega t} - 1\}. \tag{31}$$

Then substituting Eq. (30) into Eq. (29) yields

$$\left|\rho_\pm(t)\right\rangle_I = \exp\{\lambda_\pm(t)\hat{\eta}^\dagger - \lambda_\pm^*(t)\hat{\eta}\}\left|\rho_\pm(0)\right\rangle_s, \tag{32}$$

Using Eq. (8) and Eq. (10) one can decompose Eq. (32) as follows

$$\begin{aligned}\left|\rho_\pm(t)\right\rangle_I &= \exp\{\lambda_\pm(t)\hat{\eta}^\dagger - \lambda_\pm^*(t)\hat{\eta}\}\rho_\pm(0)\left|\eta=0\right\rangle, \\ &= e^{\lambda_\pm\hat{a}^\dagger - \lambda_\pm^*\hat{a}}e^{-\lambda_\pm\hat{b}+\lambda_\pm^*\hat{b}^\dagger}\rho_\pm(0)\left|\eta=0\right\rangle, \\ &= e^{\lambda_\pm\hat{a}^\dagger - \lambda_\pm^*\hat{a}}\rho_\pm(0)e^{-(\lambda_\pm\hat{a}^\dagger - \lambda_\pm^*\hat{a})}\left|\eta=0\right\rangle, \\ &= \hat{D}(\lambda_\pm)\rho_\pm(0)\hat{D}^\dagger(\lambda_\pm)\left|\eta=0\right\rangle.\end{aligned} \tag{33}$$

Where $\hat{D}(\lambda) \equiv e^{\lambda_{\pm}\hat{a}^{\dagger} - \lambda_{\pm}^{*}\hat{a}}$ is defined displaced operator of first mode, then to find density operator $\rho_{\pm}(t)$ in Schrodinger picture we should return rotational interaction picture

$$\begin{aligned} \left|\rho_{\pm}(t)\right\rangle &= e^{\frac{\gamma t}{2}\hat{L}}\left|\rho_{\pm}(t)\right\rangle_{I} \\ &= e^{\frac{\gamma t}{2}\{2\hat{a}\hat{b} - \hat{a}^{\dagger}\hat{a} - \hat{b}^{\dagger}\hat{b}\}}\left|\rho_{\pm}(t)\right\rangle_{I} \\ &= e^{\frac{\gamma t}{2}\{2\hat{a}\hat{b} - \hat{a}^{\dagger}\hat{a} - \hat{b}^{\dagger}\hat{b}\}}\hat{D}(\lambda_{\pm})\rho_{\pm}(0)\hat{D}^{\dagger}(\lambda_{\pm})\left|\eta = 0\right\rangle. \end{aligned} \quad (34)$$

To solve above equation one can immediately understand that this equation is similar in form to the master equation of lossy channel (or cavity at zero temperature) [17, 18], therefore we have

$$\rho_{\pm}(t) = e^{-\frac{\gamma t}{2}\hat{a}^{\dagger}\hat{a}}\left\{\sum_{n=0}^{\infty}\frac{T^{n}}{n!}\hat{a}^{n}\hat{D}(\lambda_{\pm})\rho_{\pm}(0)\hat{D}^{\dagger}(\lambda_{\pm})\hat{a}^{\dagger n}\right\}e^{-\frac{\gamma t}{2}\hat{a}^{\dagger}\hat{a}}. \quad (35)$$

Consequently density operator in Schrodinger picture reads as follows

$$\hat{\rho}_{\pm}^{s}(t) = \sum_{n=0}^{\infty}\frac{T^{n}}{n!}e^{-(i\omega + \frac{\gamma}{2})t\hat{a}^{\dagger}\hat{a}}\hat{a}^{n}\hat{D}(\lambda_{\pm})\hat{\rho}_{\pm}^{s}(0)\hat{D}^{\dagger}(\lambda_{\pm})\hat{a}^{\dagger n}e^{(i\omega - \frac{\gamma}{2})t\hat{a}^{\dagger}\hat{a}}. \quad (36)$$

Similarly, the explicit form of evolution of density operator $\hat{\rho}_c$ in Eq. (21) can be derived. Acting on both sides of Eq. (21) with $\left|\eta = 0\right\rangle$ we have

$$\begin{aligned} \frac{d\left|\rho_c(t)\right\rangle}{dt} &= \left\{\frac{\gamma}{2}(2\hat{a}\hat{b} - \hat{a}^{\dagger}\hat{a} - \hat{b}^{\dagger}\hat{b}) - i\lambda(\hat{a}e^{-i\omega t} + \hat{a}^{\dagger}e^{i\omega t} + \hat{b}e^{i\omega t} + \hat{b}^{\dagger}e^{-i\omega t})\right\}\left|\rho_c(t)\right\rangle \\ &= \left\{\frac{\gamma}{2}\hat{L} + \hat{g}_c(t)\right\}\left|\rho_c\right\rangle. \end{aligned} \quad (37)$$

Where

$$\begin{aligned} \hat{g}_c(t) &= -i\lambda\{(\hat{a} + \hat{b}^{\dagger})e^{-i\omega t} + (\hat{a}^{\dagger} + \hat{b})e^{i\omega t}\} \\ &= -i\lambda\{\hat{\xi}e^{-i\omega t} + \hat{\xi}^{\dagger}e^{i\omega t}\}. \end{aligned} \quad (38)$$

It is a simple task to check following commutation relations

$$\begin{array}{ll} \left[\hat{L}, \hat{\xi}\right] = 3\hat{a} - \hat{b}^{\dagger} = 2\hat{a} + \hat{\eta} = \hat{M}, & \left[\hat{L}, \hat{M}\right] = \hat{\xi}, \\ \left[\hat{L}, \hat{\xi}^{\dagger}\right] = 3\hat{b} - \hat{a}^{\dagger} = 2\hat{b} - \hat{\eta}^{\dagger} = \hat{N}, & \left[\hat{L}, \hat{N}\right] = \hat{\xi}^{\dagger}. \end{array} \quad (39)$$

To solve Eq. (37) we defined "dissipative interaction picture" for $\rho_c(t)$ as follows

$$\left|\rho_c(t)\right\rangle_I = e^{-\frac{\gamma t}{2}\hat{L}}\left|\rho_c(t)\right\rangle. \quad (40)$$

Therefore, we have

$$\frac{d\left|\rho_c(t)\right\rangle_I}{dt} = \hat{g}_c^I(t)\left|\rho_c(t)\right\rangle_I, \qquad \hat{g}_c^I(t) = e^{-\frac{\gamma t}{2}\hat{L}}\hat{g}_c(t)e^{\frac{\gamma t}{2}\hat{L}}. \quad (41)$$

Making use of commutation relations Eq. (39) one can find

$$\hat{g}_c^I(t) = -i\lambda\, e^{-\frac{\gamma t}{2}[\hat{L},\ldots]}\{\hat{\xi}e^{-i\omega t} + \hat{\xi}^\dagger e^{i\omega t}\}$$
$$= -i\lambda\left\{e^{-i\omega t}\left(\hat{\xi}\cosh\frac{\gamma t}{2} - \hat{M}\sinh\frac{\gamma t}{2}\right) + e^{i\omega t}\left(\hat{\xi}^\dagger\cosh\frac{\gamma t}{2} - \hat{N}\sinh\frac{\gamma t}{2}\right)\right\}. \tag{42}$$

Because of $\left[\hat{g}_c^I(t), \hat{g}_c^I(t')\right] \neq 0$ we can not integrate Eq. (40) thus we defined new operators $\hat{A}(t), \hat{B}(t)$ as follows

$$\hat{A}(t) = -i\lambda\cosh\frac{\gamma t}{2}(\hat{\xi}e^{-i\omega t} + \hat{\xi}^\dagger e^{i\omega t}),$$
$$\hat{B}(t) = i\lambda\sinh\frac{\gamma t}{2}(\hat{M}e^{-i\omega t} + \hat{N}e^{i\omega t}). \tag{43}$$

Now by using of appendix we can decompose Eq. (41) and find

$$\left|\rho_c(t)\right\rangle_I = \exp\left\{\int_0^t \hat{B}(s)ds\right\}\exp\left\{\int_0^t \hat{A}(s)ds\right\}\exp\left\{\int_0^t ds\int_0^s ds' f(s,s')\right\}\left|\rho_c(0)\right\rangle_I. \tag{44}$$

Where

$$f(s,s') = \left[\hat{A}(s), \hat{B}(s')\right]$$
$$= -8\lambda^2\left(\cosh(\frac{\gamma}{2}s)\sinh(\frac{\gamma}{2}s')\cos(\omega(s-s'))\right), \tag{45}$$

And

$$\int_0^t ds\int_0^s ds' f(s,s') = F(s,s'). \tag{46}$$

Therefore we have

$$\left|\rho_c(t)\right\rangle_I = e^{F(s,s')}\exp\left\{i\lambda\int_0^t \sinh(\frac{\gamma s}{2})(\hat{M}e^{-i\omega s} + \hat{N}e^{i\omega s})ds\right\}\exp\left\{-i\lambda\int_0^t \cosh(\frac{\gamma s}{2})(\hat{\xi}e^{-i\omega s} + \hat{\xi}^\dagger e^{i\omega s})ds\right\}\left|\rho_c(0)\right\rangle_I$$
$$= e^{F(s,s')}e^{\mu_2^*(t)\hat{M}-\mu_2(t)\hat{N}}e^{\mu_1(t)\hat{\xi}^\dagger-\mu_1^*(t)\hat{\xi}}\left|\rho_c(0)\right\rangle_I \tag{47}$$

Where

$$\mu_1(t) \equiv -i\lambda\int_0^t \cosh(\frac{\gamma s}{2})e^{i\omega s}ds, \qquad \mu_2(t) \equiv -i\lambda\int_0^t \sinh(\frac{\gamma s}{2})e^{i\omega s}ds. \tag{48}$$

To find the density operator $\rho_c(t)$ we decompose any of the operators $e^{\mu_1(t)\hat{\xi}^\dagger - \mu_1^*(t)\hat{\xi}}$ and $e^{\mu_2^*(t)\hat{M} - \mu_2(t)\hat{N}}$ as a product of operators, one related to the physical mode and the other to the fictitious mode, therefore we have

$$e^{\mu_1(t)\hat{\xi}^\dagger - \mu_1^*(t)\hat{\xi}}\left|\rho_c(0)\right\rangle_s = e^{\mu_1(\hat{a}^\dagger + \hat{b}) - \mu_1^*(\hat{a} + \hat{b}^\dagger)}\hat{\rho}_c(0)\left|\eta = 0\right\rangle$$
$$= e^{\mu_1\hat{a}^\dagger - \mu_1^*\hat{a}}e^{\mu_1\hat{b} - \mu_1^*\hat{b}^\dagger}\hat{\rho}_c(0)\left|\eta = 0\right\rangle$$
$$= e^{\mu_1\hat{a}^\dagger - \mu_1^*\hat{a}}\hat{\rho}_c(0)e^{\mu_1\hat{a}^\dagger - \mu_1^*\hat{a}}\left|\eta = 0\right\rangle$$
$$= \hat{D}(\mu_1)\hat{\rho}_c(0)\hat{D}^\dagger(-\mu_1)\left|\eta = 0\right\rangle \tag{49}$$

And

$$\begin{aligned}
e^{\mu_2^*\hat{M}-\mu_2\hat{N}}e^{\mu_1^*(t)\hat{\xi}-\mu_1(t)\hat{\xi}^\dagger}\left|\rho_c(0)\right\rangle_I &= e^{\mu_2^*(\hat{\eta}+2\hat{a})}e^{\mu_2(\hat{\eta}^\dagger-2\hat{b})}\hat{D}(\mu_1)\rho_c(0)\hat{D}^\dagger(-\mu_1)\left|\eta=0\right\rangle \\
&= e^{2\mu_2^*\hat{a}}e^{\mu_2^*\hat{\eta}}e^{\mu_2\hat{\eta}^\dagger}\hat{D}(\mu_1)\rho_c(0)\hat{D}^\dagger(-\mu_1)e^{-2\mu_2\hat{a}^\dagger}\left|\eta=0\right\rangle \\
&= e^{2\mu_2^*\hat{a}}e^{\mu_2^*\hat{a}+\mu_2\hat{a}^\dagger}e^{-\mu_2\hat{b}-\mu_2^*\hat{b}^\dagger}\hat{D}(\mu_1)\rho_c(0)\hat{D}^\dagger(-\mu_1)e^{-2\mu_2\hat{a}^\dagger}\left|\eta=0\right\rangle \\
&= e^{2\mu_2^*\hat{a}}e^{\mu_2^*\hat{a}+\mu_2\hat{a}^\dagger}\hat{D}(\mu_1)\rho_c(0)\hat{D}^\dagger(-\mu_1)e^{-2\mu_2\hat{a}^\dagger}e^{-\mu_2\hat{a}^\dagger-\mu_2^*\hat{a}}\left|\eta=0\right\rangle.
\end{aligned} \quad (50)$$

By means of $e^{\hat{A}+\hat{B}} = e^{\hat{A}}e^{\hat{B}}e^{-\frac{1}{2}[\hat{A},\hat{B}]}$, one can recast Eq. (50) into

$$\begin{aligned}
\left|\rho_c(t)\right\rangle_I &= e^{F(s,s')}e^{2\mu_2^*\hat{a}}e^{\mu_2^*\hat{a}+\mu_2\hat{a}^\dagger}\hat{D}(\mu_1)\hat{\rho}_c(0)\hat{D}^\dagger(-\mu_1)e^{-2\mu_2\hat{a}^\dagger}e^{-\mu_2\hat{a}^\dagger-\mu_2^*\hat{a}}\left|\eta=0\right\rangle \\
&= e^{F(s,s')-2|\mu_2|^2}e^{2\mu_2^*\hat{a}}e^{\mu_2^*\hat{a}+\mu_2\hat{a}^\dagger}\hat{D}(\mu_1)\hat{\rho}_c(0)\hat{D}^\dagger(-\mu_1)e^{-\mu_2\hat{a}^\dagger-\mu_2^*\hat{a}}e^{-2\mu_2\hat{a}^\dagger}\left|\eta=0\right\rangle \\
&= e^{F(s,s')-2|\mu_2|^2}e^{2\mu_2^*\hat{a}}e^{2\mu_2^*\hat{a}-\mu_2^*\hat{a}+\mu_2\hat{a}^\dagger}\hat{D}(\mu_1)\hat{\rho}_c(0)\hat{D}^\dagger(-\mu_1)e^{-2\mu_2\hat{a}^\dagger+\mu_2\hat{a}^\dagger-\mu_2^*\hat{a}}e^{-2\mu_2\hat{a}^\dagger}\left|\eta=0\right\rangle \\
&= e^{F(s,s')-4|\mu_2|^2}e^{4\mu_2^*\hat{a}}\hat{D}(\mu_2)\hat{D}(\mu_1)\hat{\rho}_c(0)\hat{D}^\dagger(-\mu_1)\hat{D}^\dagger(-\mu_2)e^{-4\mu_2\hat{a}^\dagger}\left|\eta=0\right\rangle \\
&= e^{F(s,s')-4|\mu_2|^2+\mu_1\mu_2^*+\mu_1^*\mu_2}e^{4\mu_2^*\hat{a}}\hat{D}(\mu_1+\mu_2)\hat{\rho}_c(0)\hat{D}^\dagger(-\mu_1-\mu_2)e^{-4\mu_2\hat{a}^\dagger}\left|\eta=0\right\rangle,
\end{aligned} \quad (51)$$

Where we have employed the following relations

$$\begin{aligned}
\hat{D}(\alpha)\hat{D}(\beta) &= \exp\left\{\frac{1}{2}(\alpha\beta^*-\alpha^*\beta)\right\}\hat{D}(\alpha+\beta), \\
\hat{D}^\dagger(\alpha)\hat{D}^\dagger(\beta) &= \exp\left\{\frac{1}{2}(\alpha\beta^*-\alpha^*\beta)\right\}\hat{D}^\dagger(\alpha+\beta).
\end{aligned} \quad (52)$$

Now using the result of the master equation lossy channel [18, 19], we can find the density operator $\hat{\rho}_c(t)$ in the Schrodinger picture

$$\hat{\rho}_c^s(t) = e^{F(s,s')+\mu_1\mu_2^*+\mu_1^*\mu_2-4|\mu_2|^2}\sum_{n=0}^{\infty}\frac{T^n}{n!}e^{-(i\omega+\frac{\gamma}{2})t\hat{a}^\dagger\hat{a}}\hat{a}^n e^{4\mu_2^*\hat{a}}\hat{D}(\mu_1+\mu_2)\hat{\rho}_c(0)\hat{D}^\dagger(-\mu_1-\mu_2)e^{-4\mu_2\hat{a}^\dagger}\hat{a}^{\dagger n}e^{-(\frac{\gamma}{2}-i\omega)t\hat{a}^\dagger\hat{a}}. \quad (53)$$

## 5. Wigner Function for Density Operator $\hat{\rho}_\pm$

In this section we find the Wigner functions of density operators Eq. (36) for the given initial state. Suppose the atom is initially in the up state $\left|\uparrow\right\rangle$ and the cavity mode is in a coherent state $\left|\alpha_0\right\rangle$, therefore the initial condition for density operator is

$$\rho_0 = \left|\alpha_0\right\rangle\left\langle\alpha_0\right|\otimes\left|\uparrow\right\rangle\left\langle\uparrow\right|. \quad (54)$$

Then substituting Eq. (54) into Eq. (19) and Eq. (20) yields

$$\rho_\pm(0) = \rho_c(0) = \left|\alpha_0\right\rangle\left\langle\alpha_0\right|. \quad (55)$$

Noticing

$$e^{-\mu\hat{a}^\dagger\hat{a}}\left|\alpha_0\right\rangle\left\langle\alpha_0\right|e^{-\mu^*\hat{a}^\dagger\hat{a}} = \exp\{|\alpha_0|^2(e^{-(\mu+\mu^*)}-1)\}\left|e^{-\mu}\alpha_0\right\rangle\left\langle e^{-\mu}\alpha_0\right|. \quad (56)$$

The equation Eq. (36) becomes

$$\hat{\rho}_\pm^s(t) = \left|e^{-(i\omega+\frac{\gamma}{2})t}(\alpha_0+\lambda_\pm)\right\rangle\left\langle e^{-(i\omega+\frac{\gamma}{2})t}(\alpha_0+\lambda_\pm)\right|. \quad (57)$$

We note that Eq. (57) shows that, if the initial condition for density operator $\rho_\pm(0)$ is the coherent state $\left|\alpha_0\right\rangle$, then at later times it continues to remain the coherent state with a time dependent complex parameter

$$\alpha_0(t) = e^{-(i\omega+\frac{\gamma}{2})t}(\alpha_0+\lambda_\pm). \quad (58)$$

Now, by inserting Eq. (31) into Eq. (58), we arrive at

$$\alpha_0(t) = \mp \frac{2i\lambda}{2i\omega + \gamma} + e^{-(i\omega + \frac{\gamma}{2})t}\left(\alpha_0 \pm \frac{2i\lambda}{2i\omega + \gamma}\right). \tag{59}$$

The Wigner function for density operator is defined as

$$W(\alpha, t) = Tr\{\hat{U}_0(\alpha)\hat{\rho}(t)\}, \tag{60}$$

Where the Wigner operator $\hat{U}_0(\alpha)$ has the form

$$\hat{U}_0(\alpha) = 2 : \exp\{-2(\alpha - \hat{a})(\alpha^* - \hat{a}^\dagger)\} : . \tag{61}$$

Inserting Eq. (57) and Eq. (60) into Eq. (59) yields

$$W_\pm(\alpha, t) = 2\exp\{-2|\alpha - \alpha_0(t)|^2\}. \tag{62}$$

Obviously, for $\gamma t \gg 1$ one can find $\alpha_0(t) = \mp \frac{2i\lambda}{2i\omega + \gamma}$, being independent of the initial state, this means that the system loses its memory. In the same method one can find $W_c(\alpha, t)$ for $\hat{\rho}_c^s(t)$.

## 6. Conclusions

In summery, for the first time we introduced a new type of the interaction picture related to dissipative processes, dissipative interaction picture in entangled state representation, and employed this picture to treat the time evolution of density operators in the J-C model with cavity damping. This technical method provides us with a fresh view of solving master equation in another problem as well. Moreover we have derived the Wigner function $W_\pm(\alpha, t)$ of density operator for special initial condition and we have shown that the system loses its memory when $\gamma t \gg 1$.

## Appendix

Here, we reproduce the proof of Eq. (44) given in [1]. Suppose that

$$\frac{d}{dt}\hat{V}(t) = \{\hat{A}(t) + \hat{B}(t)\}\hat{V}(t), \quad \hat{V}(0) = 1. \tag{63}$$

Where

$$\begin{bmatrix} \hat{A}(t), \hat{A}(t') \end{bmatrix} = \begin{bmatrix} \hat{B}(t), \hat{B}(t') \end{bmatrix} = 0 \\ \begin{bmatrix} \hat{A}(t), \hat{B}(t') \end{bmatrix} = f(t, t') \tag{64}$$

It is difficult to solve Eq. (63) because of commutation relation $\left[\hat{A}(t) + \hat{B}(t), \hat{A}(t') + \hat{B}(t')\right] \neq 0$, in order to we defined new operator $\hat{U}(t)$ as follows

$$\hat{U}(t) = \exp\{-\int_0^t \hat{B}(t')dt'\}\hat{V}(t), \\ \hat{V}(t) = \exp\{\int_0^t \hat{B}(t')dt'\}\hat{U}(t). \tag{65}$$

Therefore we have

$$\begin{aligned}\frac{d}{dt}\hat{U}(t) &= \frac{d}{dt}(\exp\{-\int_0^t \hat{B}(t')dt'\})\hat{V}(t) + \exp\{-\int_0^t \hat{B}(t')dt'\}\frac{d}{dt}\hat{V}(t) \\ &= -\hat{B}(t)\exp\{-\int_0^t \hat{B}(t')dt'\}\hat{V}(t) + \exp\{-\int_0^t \hat{B}(t')dt'\}\{\hat{A}(t) + \hat{B}(t)\}\hat{V}(t) \\ &= \exp\{-\int_0^t \hat{B}(t')dt'\}\hat{A}(t)\hat{V}(t).\end{aligned} \tag{66}$$

By substituting Eq. (65) into Eq. (66) yields

$$\begin{aligned}\frac{d}{dt}\hat{U}(t) &= \exp\{-\int_0^t \hat{B}(t')dt'\}\hat{A}(t)\exp\{\int_0^t \hat{B}(t')dt'\}\hat{U}(t)\\ &= \{\hat{A}(t) - \int_0^t [\hat{B}(t'),\hat{A}(t)]dt' + \frac{1}{2}\int_0^t [\hat{B}(t'),[\hat{B}(t'),\hat{A}(t)]] + ...\}\hat{U}(t) \quad (67)\\ &= \{\hat{A}(t) + \int_0^t f(t,t')dt'\}\hat{U}(t).\end{aligned}$$

Obviously, because of commutation relation

$$\left[\hat{A}(t) + \int_0^t f(t,s)ds, \hat{A}(t') + \int_0^{t'} f(t',s')ds'\right] = 0, \quad (68)$$

We can easily integrate Eq. (67) and find

$$\hat{V}(t) = \exp\{\int_0^t ds\hat{B}(s)\}\exp\{\int_0^t ds\hat{A}(s)\}\exp\{\int_0^t ds\int_0^s f(s,s')ds'\}. \quad (69)$$